\documentclass[preprint,12pt]{elsarticle}
\usepackage{latexsym, graphicx, epsfig, amsmath, amssymb, amsfonts}
\usepackage{amsmath}
\usepackage{amssymb}
\usepackage{graphicx}
\usepackage{subfig}
\usepackage{tabu}
\usepackage{adjustbox}
\usepackage{bbold}
\usepackage{bm}
\usepackage{textcomp}
\usepackage[thinc]{esdiff}
\usepackage[dvipsnames]{xcolor}
\usepackage{comment}

\newcommand\norm[1]{\left\lVert#1\right\rVert}

\begin{document}
\begin{frontmatter}
\title{Non-invasive Inference of Thrombus Material Properties with Physics-informed Neural Networks}
\author{Minglang Yin\textsuperscript{ab}}
\author{Xiaoning Zheng\textsuperscript{c}}
\author{Jay D. Humphrey\textsuperscript{d}}
\author{\\George Em Karniadakis\textsuperscript{c}\corref{cor1}}
\cortext[cor1]{Corresponding author: george\_karniadakis@brown.edu}
\address[a]{Center for Biomedical Engineering, Brown University, Providence, RI 02912}
\address[b]{School of Engineering, Brown University, Providence, RI 02912}
\address[c]{Division of Applied Mathematics, Brown University, Providence, RI 02912}
\address[d]{Department of Biomedical Engineering, Yale University, New Haven, CT  06520}

\begin{abstract}
We employ physics-informed neural networks (PINNs) to infer properties of biological materials using synthetic data. In particular, we successfully apply PINNs on inferring the thrombus permeability and visco-elastic modulus from thrombus deformation data, which can be described by the fourth-order Cahn-Hilliard and Navier-Stokes Equations. In PINNs, the partial differential equations are encoded into the loss function, where partial derivatives can be obtained through automatic differentiation (AD). In addition, to tackling the challenge of calculating the fourth-order derivative in the Cahn-Hilliard equation with AD, we introduce an auxiliary network along with the main neural network to approximate the second-derivative of the energy potential term. Our model can predict simultaneously unknown parameters and velocity, pressure, and deformation gradient fields by merely training with partial information among all data, i.e., phase-field and pressure measurements, and is also highly flexible in sampling within the spatio-temporal domain for data acquisition. We validate our model by numerical solutions from the spectral/\textit{hp} element method (SEM) and demonstrate its robustness by training it with noisy measurements. Our results show that PINNs can accurately infer the material properties with noisy synthetic data, and thus they have great potential for inferring these properties from experimental multi-modality and multi-fidelity data. 
\end{abstract}

\end{frontmatter}
\textbf{keywords:} Viscoelastic Porous Material, Physics-informed Neural Networks, Inverse Problem, Phase Field Model, Computational Fluids Dynamics
\newpage

\section{Introduction}
\label{sec:introduction}
Thrombus deformation and failure~\cite{rausch2017computational, rausch2016microstructurally} are important in deep vein thrombosis~\cite{weinmann1994deep, lensing1999deep,kyrle2005deep}, pulmonary embolism ~\cite{sevitt1961venous, mammen1992pathogenesis}, and atherothrombosis ~\cite{yasaka1993distribution, endo1998results}, where a key concern for a deformable thrombus is its failure and subsequent shedding of emboli, which could cause life-threatening complications under certain conditions. If we model a thrombus as a porous medium, where fibrin is loosely connected around the core area~\cite{wufsus2013hydraulic}, we can study its interaction with blood flow using mathematical models, i.e., the Cahn-Hilliard and Navier-Stokes Equations~\cite{tierra2015multicomponent}. Parameters, i.e., the permeability and visco-elastic modulus, in the governing equations play an important role in thrombus mechanics. Specifically, they can be indicators for the possibility of vessel occlusion and thromboembolism, since pieces of thrombus could be detached by the local shear stress and be transported by ambient flow to distal vessels~\cite{di2016deep}. Therefore, there is a pressing need to infer material properties from measurements, which is vital to predict thrombus shape and deformation under a variety of hemodynamic conditions and to provide an assessment of the risk of thromboembolism and other clinical consequences. Similar estimation of unknown parameters from data is also a central problem in electrocardiology and medical image reconstruction ~\cite{lucas2018using, jin2017deep, hamilton2018deep, rudy1992electrocardiographic}, geophysics~\cite{menke2018geophysical, treitel2001past, sambridge1999geophysical, yang2018application}, and many other fields~\cite{tahmasebi2016stochastic, wunsch1996ocean, bishwal2007parameter, crassidis2011optimal, lieberman2013goal}.

However, the values of permeability and visco-elasticity are patient-specific and difficult to be quantified from either experimental measurements or traditional numerical simulations using the finite element or finite volume method. A variety of numerical methods for inverse problems, i.e., Bayesian approaches~\cite{chkrebtii2016bayesian, xun2013parameter, calderhead2009accelerating}, smoothing approaches~\cite{ramsay2007parameter, liang2008parameter}, and adjoint methods~\cite{piasecki1999identification, maute2003sensitivity, nguyen2016state}, have been developed to infer PDE parameters from data.

Recent advances in solving inverse problems using deep learning techniques provide us with a promising alternative to identify PDE parameters~\cite{michoski2019solving}. In particular, physics-informed neural networks (PINNs)~\cite{raissi2019physics, raissi2020hidden} is a relatively simple framework, which encodes the information from governing equations describing physical conservation laws. Specifically, the residuals of physics equations are encoded into the loss function of the neural network as constraints such that the network outputs satisfy the PDE equations, initial, and boundary conditions. Conceptually, adding these physical constraints restrict the optimizing weights and biases in a constrained space. Unlike the traditional numerical methods, PINNs is a mesh-less framework since partial derivatives can be computed with automatic differentiation (AD) in most neural networks packages, for instance, PyTorch or TensorFlow~\cite{paszke2017automatic, abadi2016tensorflow}. As a result, the residuals of PDEs can be evaluated at random points in the spatio-temporal domain for training. Additionally, for forward problems, the training data is unpaired; PINNs does not require any data other than the spatio-temporal coordinates of training point (and the initial/boundary conditions). Successful applications of PINNs range from flow visualization~\cite{raissi2020hidden}, to high-speed flows~\cite{mao2020physics}, to stochastic PDEs~\cite{zhang2019quantifying}, to fractional PDEs~ \cite{pang2019fpinns}, and cardiac flows~\cite{kissas2020machine}, to name a few. For inverse problems with unknown parameters in PDEs, PINNs can infer even hundreds of parameters based only on measurements with a limited number of training points and without any prior knowledge on the unknown parameters~\cite{raissi2020hidden}. In PINNs, solving inverse problems follows the same workflow as forward problems only by penalizing the difference between point measurements and model predictions to the loss function. Unknown values of the parameters are set as model variables such that they can be optimized based on the gradients of the loss function with respect to their value. The potential of PINNs to infer parameters or their distributions has been explored for highly-nonlinear~\cite{raissi2019physics}, stochastic~\cite{zhang2019quantifying}, ill-posed~\cite{raissi2020hidden, raissi2019physics}, multi-fidelity problems~\cite{meng2020composite} and other cases~\cite{tartakovsky2018learning, chen2020physics}. 

In this work, we apply PINNs to identify two values of parameters, namely, the permeability and visco-elastic modulus in the Cahn-Hilliard and Navier-Stokes equation. This is perhaps the first attempt to leverage the power of PINNs as a new method to infer physiological parameters using high-order multi-physics and multi-field nonlinear PDEs. In addition, to tackling the challenge of calculating the fourth-order derivative in the Cahn-Hilliard equation with AD, we introduce an auxiliary network along with the main neural network to approximate the second-derivative of the energy potential term. Moreover, we investigate the effects of the number of training points, the influence of noisy data, and different types of data on the accuracy of our inferred results. 

The remainder of the paper is organized as follows: In section \ref{sec:methods}, we present the Cahn-Hilliard and Navier-Stokes system of equations as well as the PINN model. In section \ref{sec:results}, we present the fields construction and parameter inference results for a thrombus and a biofilm in a channel. We also explore the sensitivity of the PINN predictions by reducing the number of training data, adding noise, and using partial data from some of the fields. We conclude in section \ref{sec:discussion} with a brief summary.

\section{Methods}
\label{sec:methods}
\subsection{Cahn-Hilliard and Navier-Stokes Equations}
Mechanical interaction between thrombus and blood flow as a fluid-structure interaction (FSI) problem can be modeled by the Cahn-Hilliard and phase-field coupled Navier-Stokes equations (referred as Navier-Stokes equations) in fully-Eulerian coordinates, which are derived by minimizing the free energy of the system \cite{zheng2020three}:
\begin{align}
    \rho(\frac{\partial \mathbf{u}}{\partial t} + \mathbf{u}\cdot\nabla\mathbf{u}) +\nabla p &= \nabla \cdot (\bm{\sigma_{vis}} + \bm{\sigma_{coh}} + \bm{\sigma_{ela})} - \mu\frac{(1-\phi)\mathbf{u}}{2\kappa(\phi)},\\
    \nabla\cdot \mathbf{u} &= 0, \\
    \frac{\partial \bm{\psi}}{\partial t} + \mathbf{u}\cdot \nabla\bm{\psi} &= 0, \\
    \frac{\partial \phi}{\partial t} + \mathbf{u}\cdot \nabla \phi &= \tau \Delta \omega,\\
    \omega &= \Delta \phi + \gamma g(\phi),
\end{align}
where $\mathbf{u}(\mathbf{x}, t)$, $p(\mathbf{x}, t)$ $\bm{\sigma}(\mathbf{x}, t)$, and $\phi(\mathbf{x}, t)$ represent the velocity, pressure, stress tensor, and phase field; $g(\phi)$ equals the derivative of the double-well potential $(\phi^{2}-1)^{2}/4h^{2}$, where $h$ is the interfacial length; $\psi = [\psi_{1}, \psi_{2}]$ denotes the auxiliary vector whose gradients are the components of the deformation gradient tensor $\mathbf{F}$ as follows:
\begin{equation}
\nonumber
    \mathbf{F} := \begin{bmatrix}
            -\frac{\partial \psi_{1}}{\partial y} & -\frac{\psi_{2}}{\partial y}\\
            \frac{\partial \psi_{1}}{\partial x} & \frac{\partial \psi_{2}}{\partial x}
            \end{bmatrix}.
\end{equation}
Equation (1) is the Navier-Stokes equation with viscous, elastic, and cohesive stresses, respectively, which can be written as:
\begin{align}
    &\bm{\sigma_{vis}} = \mu \nabla u, \\
    &\bm{\sigma_{ela}} = \nabla\cdot(\lambda_{e}\frac{(1-\phi)}{2}(\mathbf{F}\mathbf{F}^{T}-\mathbf{I})),\\
    &\bm{\sigma_{coh}} = \lambda\nabla\cdot(\nabla\phi \otimes \phi).
\end{align}
Equation (2) is the continuity equation and equation (3) denotes the transport of $\psi$. The fourth-order Cahn-Hilliard equation is decoupled into two second-order equations in equations (4) and (5) for formulating the weak form; $\gamma$, $\tau$, and $\lambda$ are the interfacial mobility, relaxation parameter, and mixing energy density, respectively. Note that the quantities of interest are visco-elastic modulus $\lambda_{e}$ and permeability $\kappa(\phi)$, which are to be determined from the data by PINNs. Other PDE parameters are assumed as known.

We impose Dirichlet boundary conditions $\textbf{u} = g, (\textbf{x}, t) \in \Gamma_{i} \times (0, T)$ for velocity at the inlet $\Gamma_{i}$, and no-slip boundary on the wall $\Gamma_{w}$. Neumann boundary conditions, i.e., $\frac{\partial \phi}{\partial \textbf{n}} = \frac{\partial \omega}{\partial \textbf{n}} = \frac{\partial \bm{\psi} }{\partial \textbf{n}} = 0, \textbf{x} \in \Gamma_{w} \cup \Gamma_{i} \cup \Gamma_{o}$ are imposed for $\psi$, $\phi$, and $\omega$ on all boundaries, and for pressure and velocity at the outlet $\Gamma_{o}$. This model is feasible for both 2D and 3D but we only consider two-dimensional (2D) physical domain in this paper for proof of concept demonstration.

\subsection{Physics-Informed Neural Networks (PINNs)}
\begin{figure}
    \centering
    \includegraphics[width=1\textwidth]{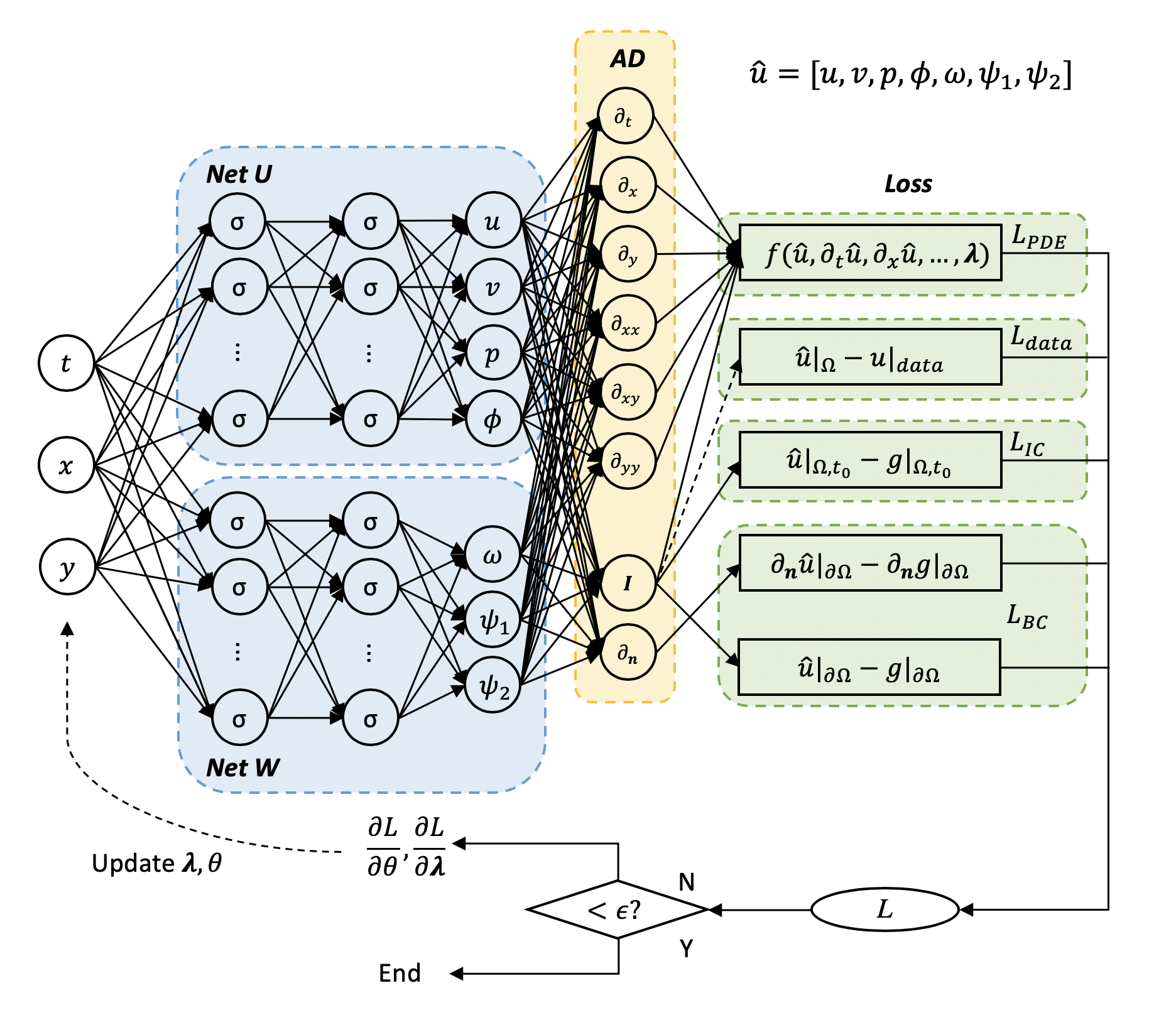}
    \caption{\textbf{Schematic of a PINN for solving inverse problem for Cahn-Hilliard and  Navier-Stokes  PDEs.} Circled by blue boxes, \textit{Net U} and \textit{Net W} represent surrogate models for the PDEs solution whose derivatives can be computed with automatic differentiation (AD). The computed derivatives are used in the loss function to restrict model outputs such that they satisfy the system of PDEs in $\Omega$. For inverse problems, the residual between sensor measurements $u|_{data}$ and model outputs $\hat{u}_{\Omega}$ are included in the loss function. We use ADAM to optimize the model parameters $\bm{\theta}$ (weights and biases) and search the unknown values of the material parameters from the PDEs $\bm{\lambda}$ to minimize the loss function.}
    \label{fig:net_struct}
\end{figure}

In Fig. \ref{fig:net_struct} we show a schematic of PINNs. Given the time \textit{t} and coordinates \textit{x, y} of training points as inputs, we construct two fully-connected neural networks, \textit{Net U} and \textit{Net W}, where the outputs of \textit{Net U} represents a surrogate model for the PDE solutions \textit{u, v, p}, and \textit{$\phi$} and the outputs of \textit{Net W} are PDE solutions \textit{$\omega$, $\psi_{1}$}, and \textit{$\psi_{2}$}. We denote the PDE solutions as $\hat{u}$ concatenated by the outputs from \textit{Net U} and \textit{Net W}, whose derivatives with respect to the inputs are calculated using AD. Then, we formulate the total loss \textit{L} as the combination of PDEs residual loss (\textit{$L_{PDE}$}), initial and boundary condition loss (\textit{$L_{IC}$}, \textit{$L_{BC}$}), and data loss \textit{$L_{data}$} as follows:
\begin{equation}
    L = \omega_{1}L_{PDE} + \omega_{2}L_{IC} + \omega_{3}L_{BC} + \omega_{4}L_{data},
\end{equation}
and
\begin{align}
    L_{PDE}(\theta, \bm{\lambda}; X_{PDE}) &= \frac{1}{ \left| X_{PDE} \right| }  \sum\limits_{\textbf{x}\in X_{PDE}} \norm{f(\textbf{x}, \partial_{t}{\hat{\textbf{u}}}, \partial_{x}\hat{\textbf{u}},...,\partial_{xx}{\hat{\textbf{u}}},...;\bm{\lambda})}_{2}^{2},\\
    L_{BC}(\theta, \bm{\lambda}; X_{BC}) &= \frac{1}{\left| X_{BC} \right|} \sum\limits_{\textbf{x}\in X_{BC}}\norm{\mathfrak{B(\hat{\textbf{u}}, \textbf{x})}}_{2}^{2},\\
    L_{IC}(\theta, \bm{\lambda}; X_{IC}) &= \frac{1}{\left| X_{IC} \right|} \sum\limits_{\textbf{x}\in X_{IC}}\norm{\hat{\textbf{u}} - \textbf{u}_{t_{0}}}_{2}^{2}, \\
    L_{data}(\theta, \bm{\lambda}; X_{data}) &= \frac{1}{\left| X_{data} \right|} \sum\limits_{\textbf{x}\in X_{data}}\norm{\hat{\textbf{u}} - \textbf{u}_{data}}_{2}^{2},
\end{align}
where $\omega_{1}$, $\omega_{2}$, $\omega_{3}$, and $\omega_{4}$ are the weights of each term. The training sets $X_{PDE}$, $X_{BC}$, and $X_{IC}$ are sampled from the inner spatio-temporal domain, boundaries, and initial snapshot, respectively. $X_{data}$ is the set that contains sensor coordinates and point measurements; $\left| \cdot \right|$ denotes the number of training data in the training set. In particular, $\mathfrak{B}$ represents a combination of the Dirichlet and Neumann residuals at boundaries. Finally, we optimize the model parameters $\bm{\theta}$ and the PDE parameters $\bm{\lambda}=[\lambda_{e}, \kappa]$ by minimizing the total loss $L(\bm{\theta}, \bm{\lambda})$ iteratively until the loss satisfies the stopping criteria. Optimizing the total loss is a searching process for $\bm{\lambda}$ such that the outputs of the PINN satisfy the PDE system, initial/boundary conditions, and point measurements. We use the mean relative $L_{2}$ error ($\epsilon$), same as in \cite{raissi2020hidden}, to quantify errors between reference data and model predictions:
\begin{equation}
    \epsilon := (\frac{1}{N}\sum_{i}^{N}[\hat{u}(\textbf{x}_{i})-u(\textbf{x}_{i})]^{2})/(\frac{1}{N}\sum_{i}^{N}[u(\textbf{x}_{i})-\frac{1}{N}\sum_{i}^{N}u(\textbf{x}_{i})]^{2})
\end{equation}

\section{Results}
\label{sec:results}
To demonstrate the inference ability of the PINN model, we adopt four representative cases for parameters inference. The high-resolution training datasets are generated from the spectral/$hp$ element solver $\mathcal{NEKTAR}$ \cite{karniadakis2013spectral} coupled with the Cahn-Hilliard equations with 3$rd$-order Jacobi polynomials. For the neural network architecture, our preliminary results suggested that using 9 hidden layers with 20 neurons per layer for \textit{Net U} and \textit{Net W} could be a good balance between the network representation capacity and the computational costs. We use the ADAM optimizer~\cite{kingma2014adam} with learning rate 0.001 to train the model for a number of epochs, which is defined as the number of complete passes through the full training dataset.

\subsection{Inference of Permeability}
\subsubsection{Thrombus in a channel with uniform permeability}
\begin{figure}
    \centering
    \includegraphics[width=1\textwidth]{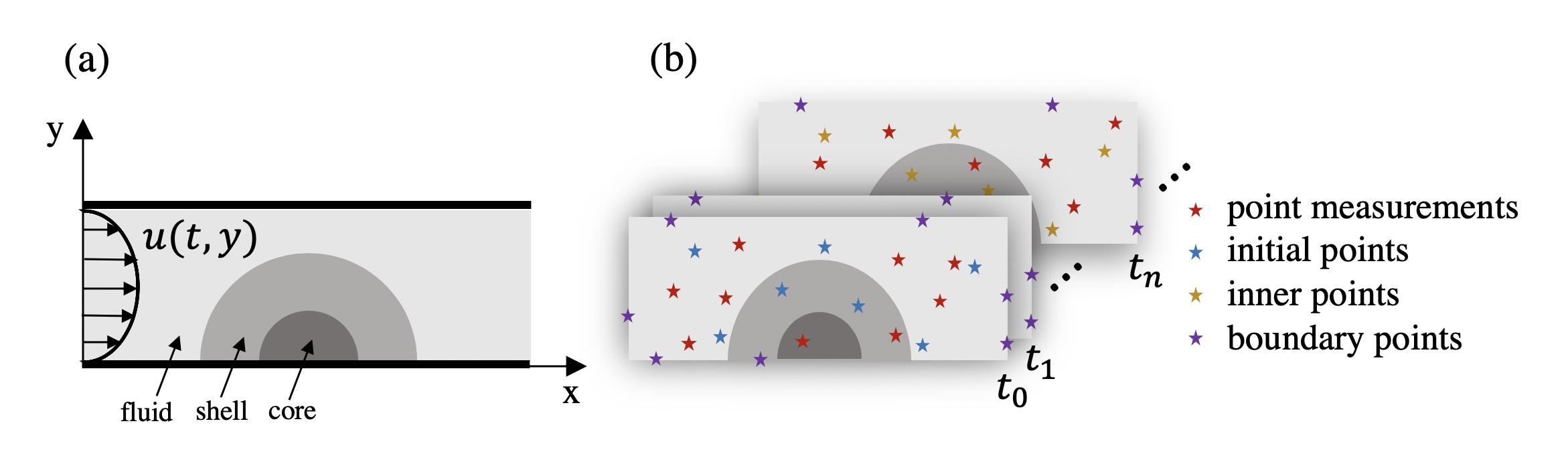}
    \caption{\textbf{2D flow past a thrombus.} (a) The computational domain is a  channel with walls on the top and bottom boundaries and the inlet flow $u(t,y)$ entering from the left side.; $\phi$=1 corresponds to fluid. A thrombus with a permeable core $\phi=-1$ and shell $\phi=0$ is present at the bottom boundary. (b) Sampling points for inferring permeability include initial points (\textcolor{ProcessBlue}{$\star$}) at the time $t_{0}$, inner points (\textcolor{Dandelion}{$\star$}) from $t_{1}$ to $t_{n}$, boundary points (\textcolor{Plum}{$\star$}) on boundaries, and point measurements (\textcolor{BrickRed}{$\star$}) with PDE solutions.}
    \label{fig:K_sampling}
\end{figure}
To infer the permeability $\kappa$ in the Cahn-Hilliard and Navier-Stokes equations, we perform simulations for a semi-circle permeable thrombus in a channel with a steady parabolic flow coming from the left. We impose the Neumann type boundary condition for $\phi$, $\psi$ as $\nabla\phi\cdot\textbf{n} = \nabla\psi\cdot\textbf{n}=0$, where \textbf{n} is the unit vector perpendicular to the boundaries. We set the density $\rho$ = 1, viscosity $\mu$ = 0.1, $\lambda$ = $4.2428\times10^{-5}$, $\tau$ = $10^{-6}$, visco-elastic modulus $\lambda_{e}$ = 0, and the interface length $h$= 0.05. These parameters in PINNs are non-dimensionalized numbers so as to be consistent with the CFD solver. The thrombus is present in the middle of the channel as shown in Fig. 2(a) with a uniform permeability in the core ($\phi=-1$) and in the outer shell layer ($\phi=0$). In general, the inlet velocity $u(t, y)$ can be time-dependent flow, but in this case it is set as steady $0.3(y-2)y$. In plot (b),  we sample coordinates of training data in the initial snapshot $t_{0}$ (\textcolor{ProcessBlue}{$\star$}), inner spatio-temporal domain from $t_{1}$ to $t_{n}$ (\textcolor{Dandelion}{$\star$}), and at boundaries (\textcolor{Plum}{$\star$}). Moreover, we also sample point measurements (\textcolor{BrickRed}{$\star$}) including their coordinates and PDEs solutions in the spatio-temporal domain to calculate the data loss term in the total loss. In this section, the points measurements only contain scalar data from the phase field $\phi$ as the data source to recover the velocity field and the missing parameter $\kappa$.

\begin{figure}
    \centering
    \includegraphics[width=0.95\textwidth]{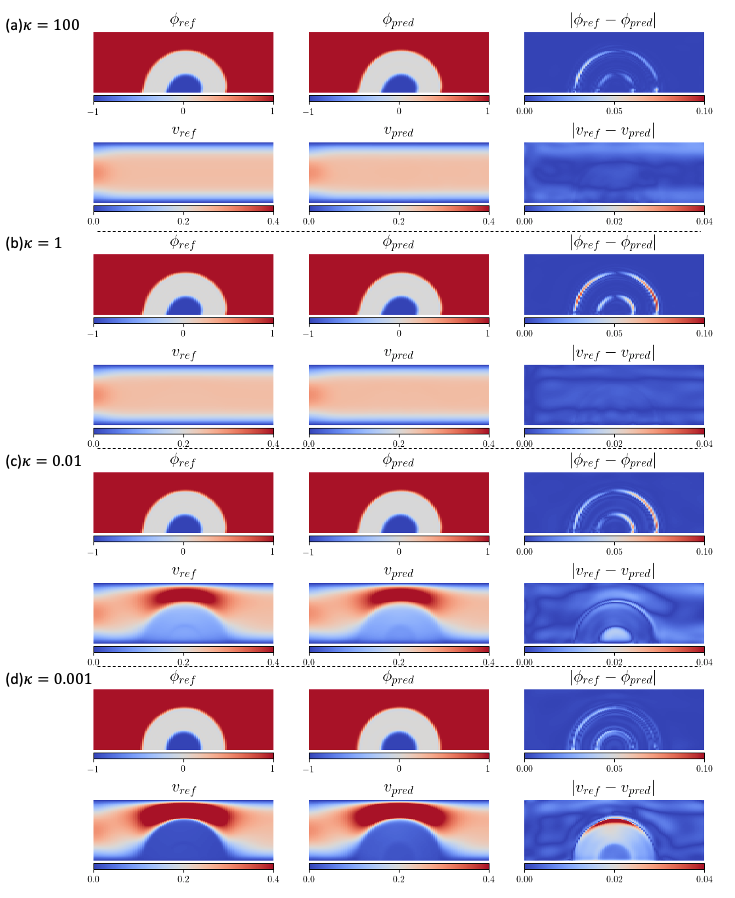}
    \caption{\textbf{Prediction and error of 2D flow past a thrombus for various permeability values at $t=0.3$.} Representative snapshots of the reference data for permeability value $\kappa$ (a) 100 (b) 1 (c) 0.01 (d) 0.001 are shown against the predicted phase and velocity field. The first column shows the reference simulation results while the second shows the results from the PINNs. The third column shows the absolute value of the difference between references and model predictions. 30\% of the phase field data are used in the training process.}
    \label{fig:K_comp}
\end{figure}

In Fig. \ref{fig:K_comp}, we first show four simulation cases with permeability values from $10^{-3}$ to $10^2$. We first consider a simple case with the value of $\kappa$ uniform across core and shell areas as proof of concept. We use 30,000 training points of the phase field to infer $\kappa$ for each case. The first column shows the reference data of phase and velocity fields at $t=0.3$ as ground truth. As shown in Fig. \ref{fig:K_comp}(a), the thrombus is permeable and fluid can penetrate the core and the shell of the thrombus. In Fig. \ref{fig:K_comp}(b)-(d), the fraction of fluid in the thrombus falls off dramatically as a result of decrease of the thrombus permeability and an accelerated area is formed on the top of the thrombus. Unlike the thrombus with large $\kappa$, fluid can hardly flow through the impermeable thrombus and the phase deformation becomes relatively small. In other words, the deformation of phase fields is noticeable when $\kappa$ is large, and becomes hardly noticeable for permeability 0.001. In Fig. \ref{fig:K_comp}, the second column depicts the predicted phase and velocity fields, and the third column shows the absolute error. The model predictions are in excellent agreement with the reference data for both phase and velocity fields. The error in the phase field is mainly distributed on the thrombus interface with the maximum absolute error smaller than 10\% for all cases. However, the velocity error grows with the decrease of permeability with the largest local error reaching 10\% under the bottleneck region. Overall, the results show that the PINN model can regress PDE solution fields and infer parameters accurately from synthetic data.

\begin{figure}
    \centering
    \includegraphics[width=1\textwidth]{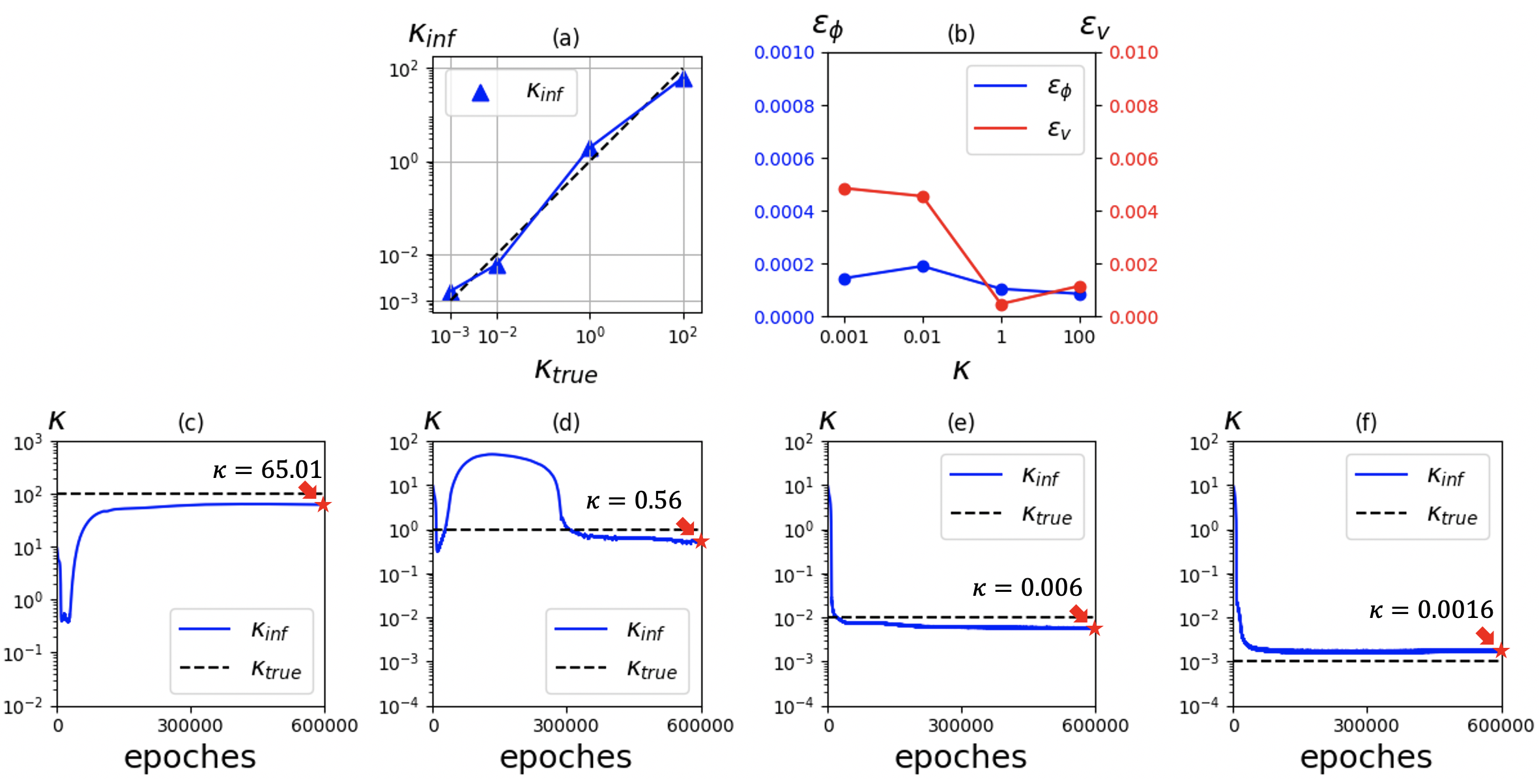}
    \caption{\textbf{Performance and history of PINN on predicting permeability of 2D flow past a thrombus with $\kappa$ ranging from 0.001 to 100.} (a) Comparison of the inferred values with the reference values. (b) Mean relative $L_{2}$ errors for phase and velocity field when $\kappa$ varies from 0.001 to 100. The plots in (c, d, e, f) illustrate the inference of the permeability with respect to the number of iterations  of PINN when $\kappa=100$, $\kappa=1$, $\kappa=0.01$, and $\kappa=0.001$. }
    \label{fig:K_vary}
\end{figure}

We summarize the results of parameters inference, mean relative $L_{2}$ errors as well as the history of inverse findings in Fig. \ref{fig:K_vary}. Plot (a) shows that all the inferred $\kappa$ values fall near the diagonal line, indicating a good agreement between the reference values and the inferred values. Plot (b) shows the mean relative $L_{2}$ error for each case: the maximum relative error for velocity prediction is below 0.6\% and for the phase field is less than 0.02\%. Figs. \ref{fig:K_vary} (c-f) depict the convergence history of parameter retrieval in the training process to the true different values of $\kappa$.

\begin{figure}
    \centering
    \includegraphics[width=1\textwidth]{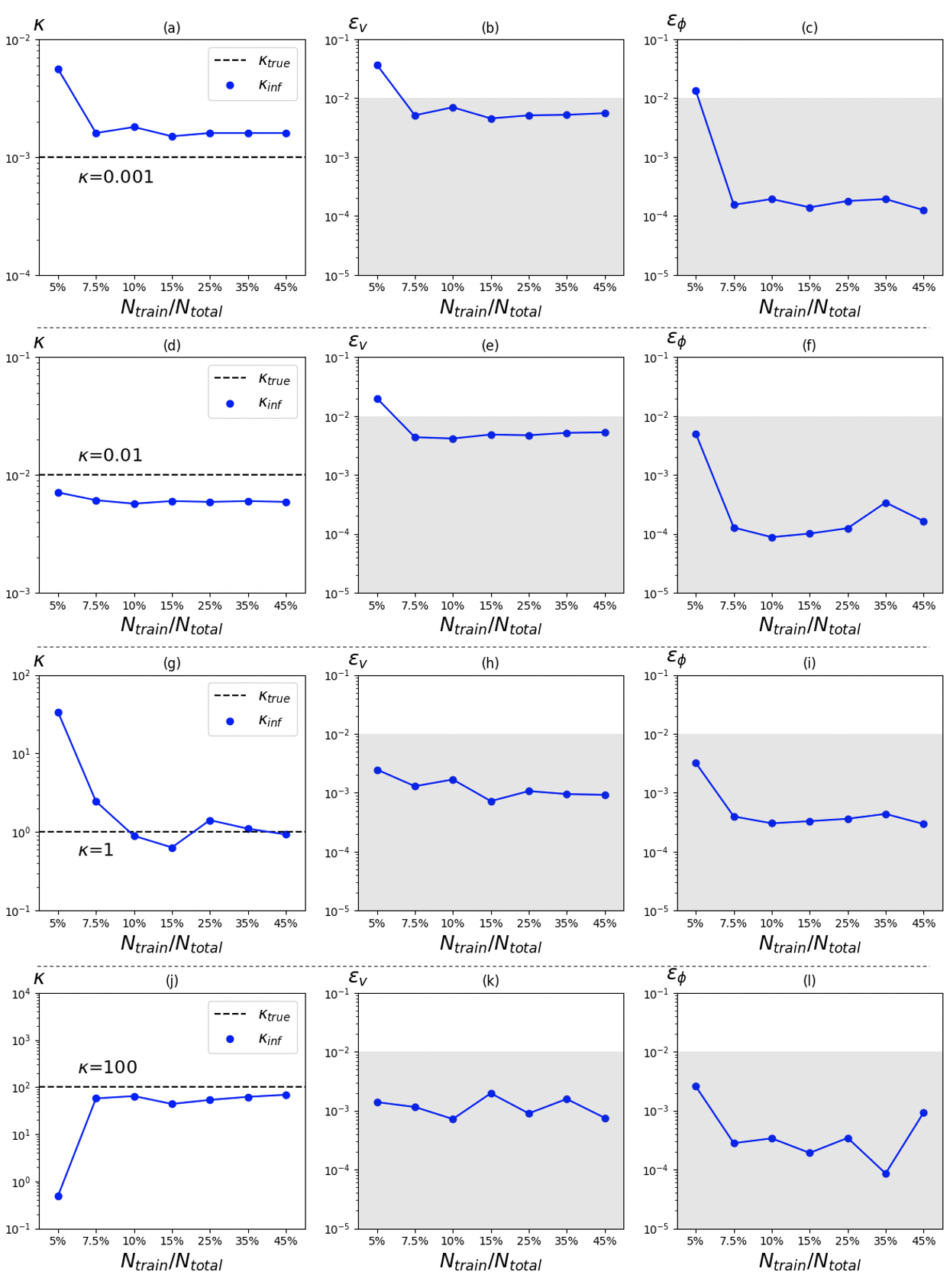}
    \caption{\textbf{Effect of the number of training points on the accuracy of PINNs.} $N_{total}$ = 200,000. Plots (a, d, g, j) show comparison of reference and inferred permeability for a different number of training points from 5\% of the total number of points to 45\%. The second and the third column show the mean of relative $L_{2}$ error for (b, e, h, k) velocity field and (c, f, i, l) phase field over time for different numbers of training points. We shaded the area where the error is lower than 1\%. }
    \label{fig:K_num}
\end{figure}

To investigate the effect of the number of training points on the model prediction ability, we retrain the model for the same cases as in Fig. \ref{fig:K_comp} and Fig. \ref{fig:K_vary} with a different number of training points from 5\% up to 45\% of the total amount of data (200,000). Figs. \ref{fig:K_num} (a, d, g, j) show the trend of the inferred values changing with the spatio-temporal resolution of the training data. Generally, the inference results converge toward the true value with mild deviations if the use of training data is greater than 7.5\% (i.e., 15,000 training points) among the total number of points. The second and the third column show the mean of relative $L_{2}$ errors for velocity and phase fields. The velocity errors for $\kappa$ equals 1 and 100 are one order smaller than those of small permeability, indicating better prediction results when $\kappa$ is large. The phase field errors are all lower than 1\% if the training data is above 7.5\% of the total number of points. Hence, for this problem, we conclude that it is sufficient to guarantee convergence and good results if 7.5\% of points are used to make inferences. 

\begin{figure}
    \centering
    \includegraphics[width=1\textwidth]{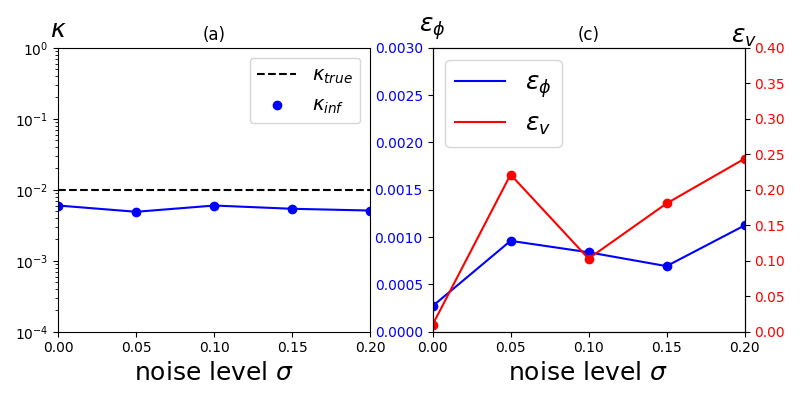}
    \caption{\textbf{Predictions of 2D flow past a thrombus trained with noisy phase field measurements.} (a) Inferred permeability and (c) mean relative $L_{2}$ errors between the model predictions and the reference for phase $\epsilon_{\phi}$ and velocity field $\epsilon_{v}$. Noise is added to the phase field with the noise level ranged from 0 to 20\%. Here, 10,000 data points are scattered in the spatio-temporal domain as the training data to infer the permeability.}
    \label{fig:K_noise}
\end{figure}

Furthermore, to validate the robustness of our model to noisy measurements, we add white noise $\mathcal{N}(0, 1)$ to the input data, i.e., phase field $\phi$, in the following way:
\begin{equation}
    \hat{\phi} = \textrm{clip}(\phi + \sigma\mathcal{N}(0, 1)), \hat{\phi}\in[-1, 1],
    \label{equ:noise}
\end{equation}
where $\sigma$ is the noise level. We add normal-distributed white noise signals to the reference data $\phi$ and impose a clip function to restrain the value of $\phi$ within -1 and 1. We test the value of the noise level $\sigma$ up to 20\% of the variance and train the model with these noisy data. The reference permeability values in these cases are set as 0.01. Fig. \ref{fig:K_noise} (a) plots the inferred $\kappa$ against various noise levels, showing that the inferred value is not affected by the added noise. In plot (b), the relative $L_{2}$ errors for phase and velocity field show slight increase as the noise level increases. In particular, the predictions of the velocity field is more sensitive to the noise as $\epsilon_{v}$ increases from less than 1\% at $\sigma=0$ to 25\% at $\sigma=0.20$. In general, the good agreement in parameter inference and small increment in fields predictions demonstrate strong robustness of the PINN model to noisy data.

\subsubsection{Thrombus in a channel with space-dependent permeability}

\begin{figure}
    \centering
    \includegraphics[width=1\textwidth]{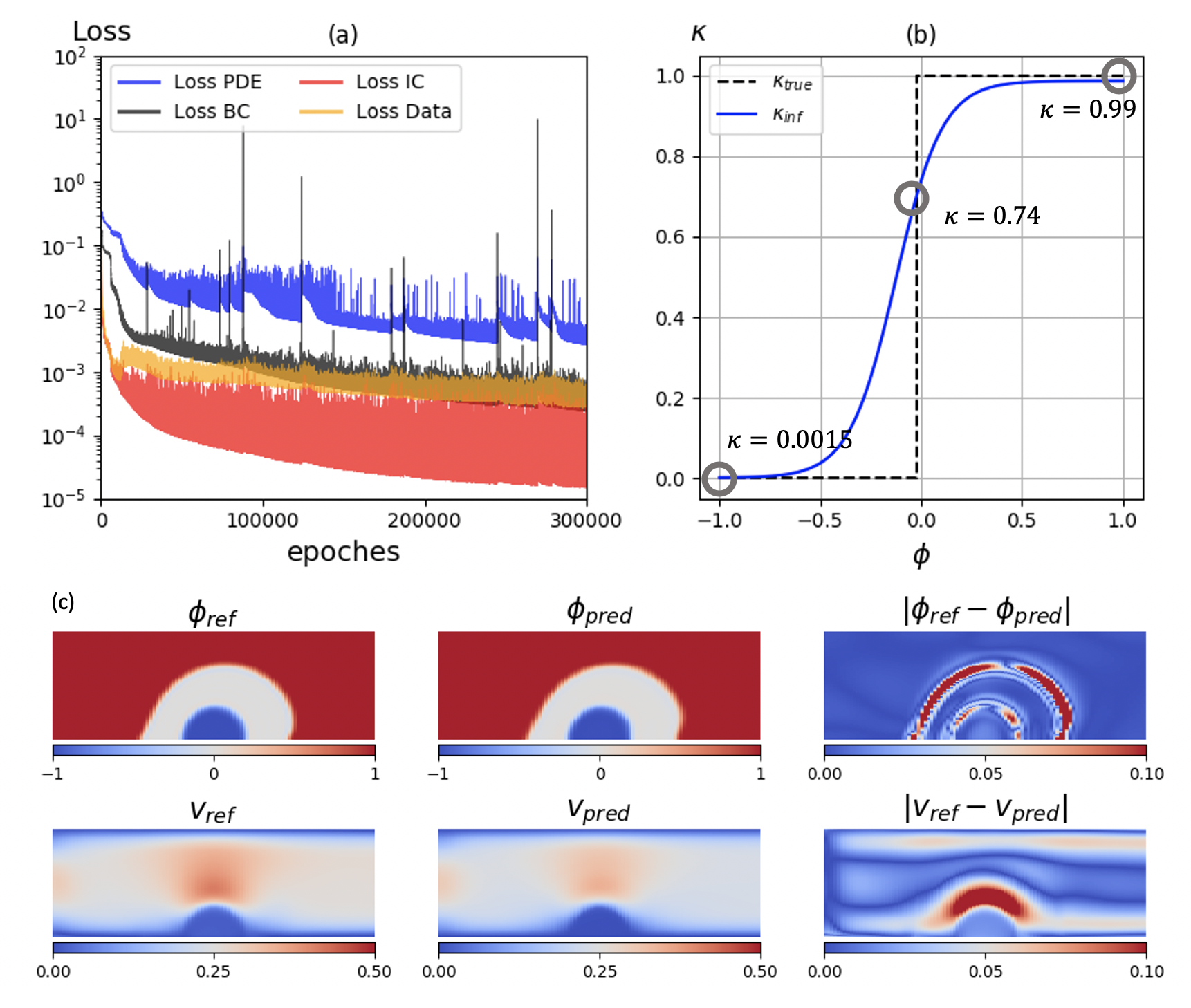}
    \caption{\textbf{2D flow past a thrombus with phase dependent permeability.} (a) History of network losses (Loss PDE, Loss IC, Loss BC, and Loss Data) and (b) inferred the permeability $\kappa$ as a function of $\phi$. (c) Comparison of phase field and velocity field for $\kappa(\phi)$ at $t$ = 0.3 and their absolute error. The core permeability is 0.001 and the shell permeability is set 1 as the actual values. 50,000 data points are scattered in the spatio-temporal domain among 30 snapshots as the training data to infer the permeability.}
    \label{fig:K_phi}
\end{figure}

Unlike the aforementioned idealistic case, in a real thrombus, the permeability varies spatially depending on the volume fraction ($\phi$), with the core area much less permeable than the outer shell. To validate the inference ability of the PINN for a space-dependent permeability, we test another case with the $\kappa=1$ for the shell area and $\kappa=0.001$ for the core area as shown in Fig. \ref{fig:K_phi}. Since the core area is hardly permeable while the shell has a larger $\kappa$, we expect to observe a non-uniform displacement from the thrombus core and shell as the outer layer moves with ambient flow and the inner layer stays still. To express such spatial variation explicitly, we utilize an equation to express the relation between $\phi$ and $\kappa$ in this case:
\begin{equation}
    \kappa(\phi) = a\tanh{b\phi + c} + d,
\end{equation}
where $a, b, c$ and $d$ are model parameters to be optimized in the PINN model. 

In Fig. \ref{fig:K_phi} we present the history of the separated term of the loss, namely PDE loss, boundary condition loss (Loss BC), initial condition loss (Loss IC), and data loss (Loss Data) in (a). Plot (b) shows the inference result for $\kappa$ as a function of $\phi$; the predicted $\kappa$ for the fluid ($\kappa(\phi=1)=0.0015$) and the core ($\kappa(\phi=-1)=0.99$) match the reference values very well while there exists a difference for the shell area ($\kappa(\phi = 0)=0.74$) since the true value equals 1. We present the reference data $\phi_{ref}$, $v_{ref}$ and the model predictions $\phi_{pred}$, $v_{pred}$ and their difference at $t=0.6$ in plot (c). The errors for the phase field are mainly distributed in and around the outlet layer of the thrombus, and the errors in velocity field are mainly confined within the shell layer. Such inconsistency in the phase field and the velocity field may be induced by the under-predicted value in the shell layer permeability, see plot  \ref{fig:K_phi} (c).

\subsection{Inference of Visco-elastic Modulus}
The visco-elastic modulus is another important physiological parameter that has to be estimated indirectly. There are few rheometry experiments to measure the visco-elastic shear modulus $\lambda_{e}$ with oscillatory shear deformation. We assume the homogeneity and isotropy of the thrombus for simplicity. To explore the viability of parameter inference from imaging data, we consider two typical setups as illustration: a thrombus in a cavity, and a biofilm in a channel.

\subsubsection{Visco-elastic thrombus in a cavity}
\begin{figure}
    \centering
    \includegraphics[width=1\textwidth]{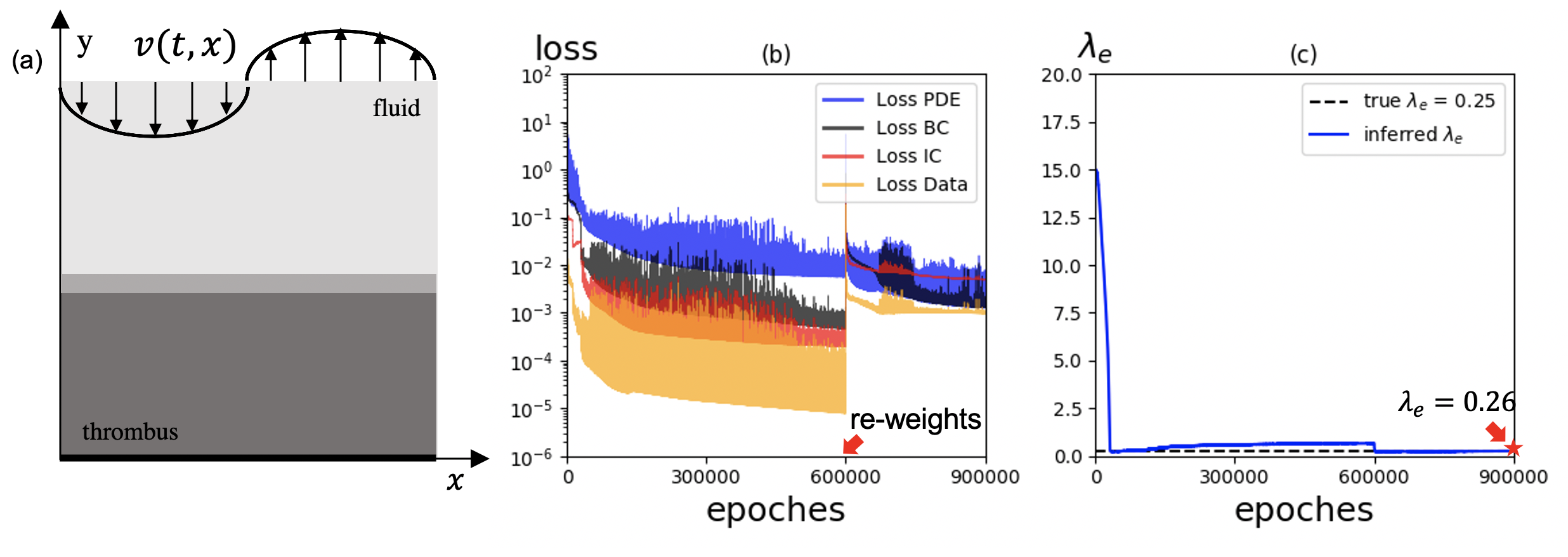}
    \caption{\textbf{(a) Schematic for visco-elastic thrombus in a cavity, (b) history losses for each term, and (c) inference of $\lambda_{e}$.} (a) We impose a time-dependent sinusoidal vertical velocity $v(t, x) =-(1-\cos(2\pi t))\sin(2\pi x)$ at the top boundary, and set the left and right sides as periodic. We changed the weight for each loss term at epoch 600,000. 30,000 training data points are scattered among the spatio-temporal domain to train the network. The training data only contain the phase field from the inner points and pressure measurements at the boundaries. \textcolor{blue}{Loss PDE}: loss for the PDEs residuals, \textcolor{black}{Loss BC}: loss for boundary conditions, \textcolor{red}{Loss IC}: loss for initial conditions, \textcolor{orange}{Loss Data}: loss for measurements data.}
    \label{fig:lmd_mem_his}
\end{figure}

We first consider a visco-elastic thrombus in a 1$\times$1 cavity as shown in Fig. \ref{fig:lmd_mem_his} (a). The top layer with the light gray color denotes the fluid phase while the bottom layer with the darker color indicates the initial state of visco-elastic thrombus. We impose a time-dependent sinusoidal vertical flow $v(t, x) = -(1-\cos(2\pi t))\sin(2\pi x)$ at the top boundary. At the left and right boundary, we set the Dirichlet boundary $\phi(y) = \tanh(y-0.5)/\sqrt{2}h$ and the periodic boundary for velocity; we also set $\nabla\phi\cdot \textbf{n}=0$ at the top boundary and $\phi = -1$ on the bottom wall. We sample 30,000 training points from 20 consecutive snapshots, each containing 10,000 points from $t$ = 0.03 to $t$ = 0.63. To train this neural network, we only utilize the phase field information and some pressure measurements at the boundaries as data sources. Such data acquisition does not require information other than the phase field from the inner domain, such as the pressure or auxiliary vector field $\psi$, and hence it can potentially be used in a real experimental setup. The weights are chosen as followed:
\begin{align}
    \omega_{1} = \omega_{3} = 1, \omega_{2} = \omega_{4} = 5, &\text{ epoch $\in$ [1, 600,000]},\\
    \omega_{1} = 10, \omega_{3} = \omega_{2} = \omega_{4} = 1, & \text{ epoch $\in$ [600,001, 900,000]},
\end{align}
We set $\rho= 1$, $\mu = 0.01$, $h = 0.02$, $\lambda = = 2.5\times 10^{-9}$, and $\tau = 10^{-4}$.

In Fig. \ref{fig:lmd_mem_his} We present in plot (b) the history of the loss for each term and in plot (c) the inferred value of the visco-elastic modulus. In plot (b), the PDE loss (\textcolor{blue}{blue line}) converges around $10^{-3}$ and the other losses balanced at the same order with the PDE loss after redistributing the weights at epoch 600,000. Another result of changing the weights is that the inferred value for $\lambda_{e}$ converges closer to the reference value 0.25. 

\begin{figure}
    \centering
    \includegraphics[width=1\textwidth]{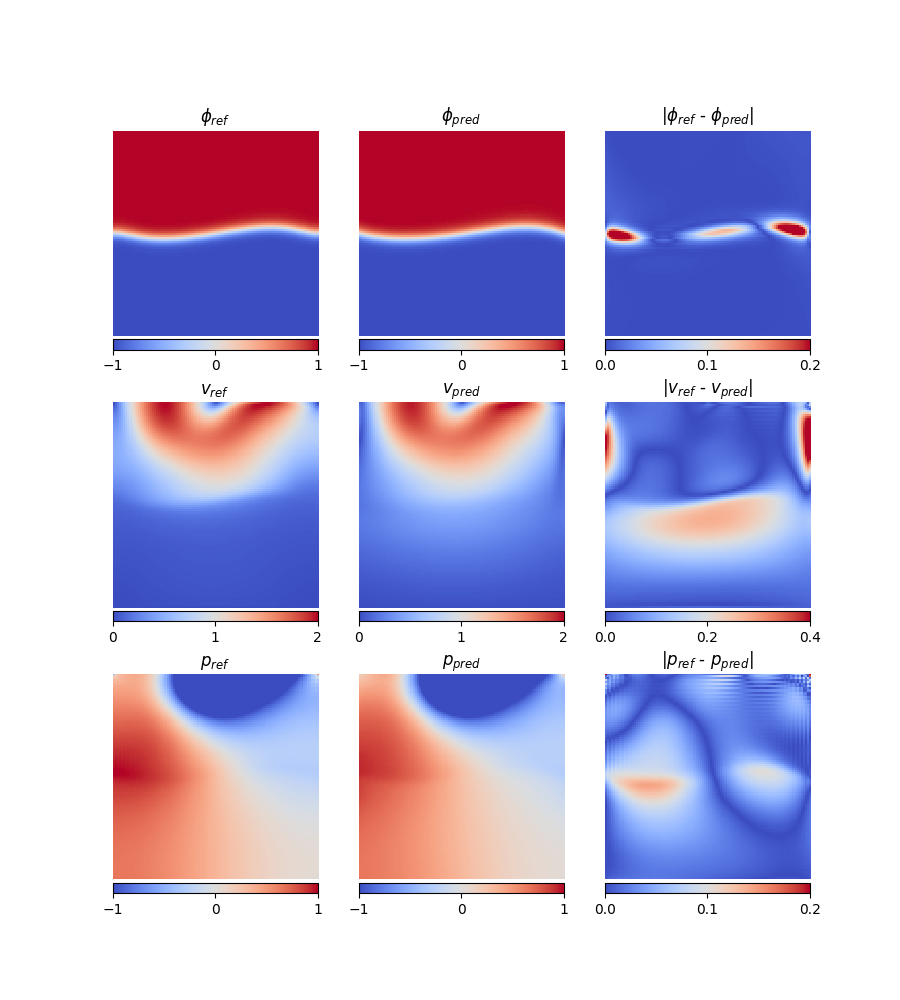}
    \caption{\textbf{Visco-elastic thrombus in a cavity.} The first column presents the phase field distribution $\phi$, velocity field $v$, and pressure distribution $p$ from reference data, respectively. The second column shows the same field predictions from the model at $t$ = 0.48. The absolute difference between the data and the model predictions are plotted in the third column.}
    \label{fig:lmd_mem}
\end{figure}

\begin{table}
\centering
\begin{adjustbox}{width=1\textwidth}
 \begin{tabular}{c | c | c | c | c | c } 
 \hline
 & $\epsilon_{v}$ & $\epsilon_{\phi}$ & $\epsilon_{p}$ & $\epsilon_{\psi_{1}}$ & $\epsilon_{\psi_{2}}$\\ 
 \hline
 \hline
 Full& $3.511\times10^{-2}$ & $1.157\times 10^{-3}$ & $2.391\times 10^{-3}$ & $2.636\times 10^{-2}$ & $2.450\times 10^{-2}$   \\
 \hline
 Half& $4.494\times10^{-3}$ & $4.129\times 10^{-4}$ & $1.057\times 10^{-3}$ & $2.516\times 10^{-2}$ & $2.236\times 10^{-2}$   \\
 \hline
\end{tabular}
\end{adjustbox}
\caption{\textbf{Summary of the mean relative $L_{2}$ error for the thrombus in a cavity problem over half and full time window.}}
\label{table:mem_err}
\end{table}

Fig. \ref{fig:lmd_mem} compares the reference data and the model predictions at time $t$ = 0.48. Phase, velocity, and pressure fields are plotted respectively on each row, and the last column plots the absolute difference between the data and predictions. We can observe excellent inferred results for $\phi$, $p$, and $v$ with some minor discrepancies at the interface layer and top periodic layer. Additionally, our model renders high-resolution results in fields construction as can be seen in the summary of the mean relative $L_{2}$ error in Table \ref{table:mem_err}. The first row shows the mean errors for each field over the full snapshots, while the second row lists the mean errors over the first half among all snapshots. As the data indicate, the errors increase as the system is developing since the full time window errors are all greater than that at the first half time window. But overall, we can conclude that the model infers the fields for each variable with satisfactory accuracy.


\subsubsection{Biofilm in a channel}

\begin{figure}
    \centering
    \includegraphics[width=1\textwidth]{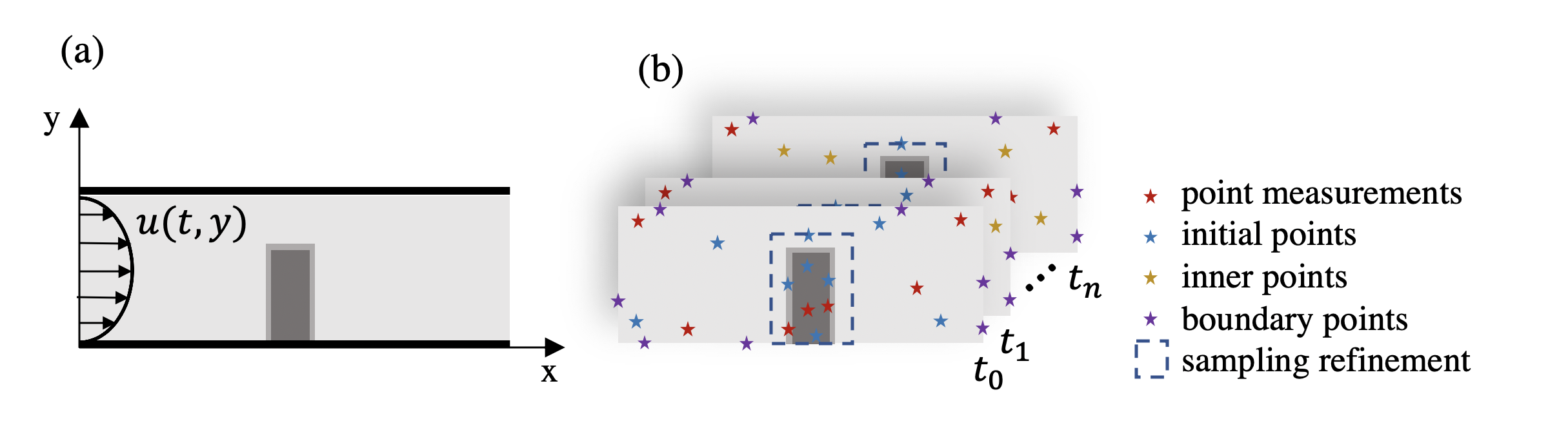}
    \caption{\textbf{2D flow past a visco-elastic biofilm setup.} (a) The computational domain is a channel with wall boundaries on the top and bottom sides, and with flow $u(t)$ entering from the left side. A visco-elastic biofilm is present at the bottom boundary. (b) Training points include initial points (\textcolor{ProcessBlue}{$\star$}) at time $t_{0}$, inner points (\textcolor{Dandelion}{$\star$}) from $t_{1}$ to $t_{n}$, boundary points (\textcolor{Plum}{$\star$}) on the boundaries, and point measurements (\textcolor{BrickRed}{$\star$}) with PDE solutions. The sampling points are refined within the dashed line area to improve the results of the PINN model.}
    \label{fig:lmd_sampling}
\end{figure}

In this section, we consider another possible scenario, where a thin visco-elastic biofilm is present in the middle of a channel with oscillatory flow $u(t, y) = 0.9\sin(2\pi t)(2y-y^{2})$ coming from the left side of the domain. We expect to observe a swinging movement of the biofilm with the oscillatory flow. Similar as the sampling strategy on the inference of permeability, we sample the four types of points from this domain as shown in Fig. \ref{fig:lmd_sampling}(b): initial points (\textcolor{ProcessBlue}{$\star$}), inner points (\textcolor{Dandelion}{$\star$}), boundary points (\textcolor{Plum}{$\star$}), and points with measurements (\textcolor{BrickRed}{$\star$}). Since the dynamics is rich in the area indicated by the black dash line, we refine the density of sampling points from $t_{0}$ to $t_{n}$ within the box for better accuracy. For this 2D flow, we set $\lambda$ = $4.2428\times10^{-5}$, $\tau$ = 0.5, $\rho$ = 1, and $\mu$ = 0.1, and the interface width $h$ = 0.04.

\begin{figure}
    \centering
    \includegraphics[width=1\textwidth]{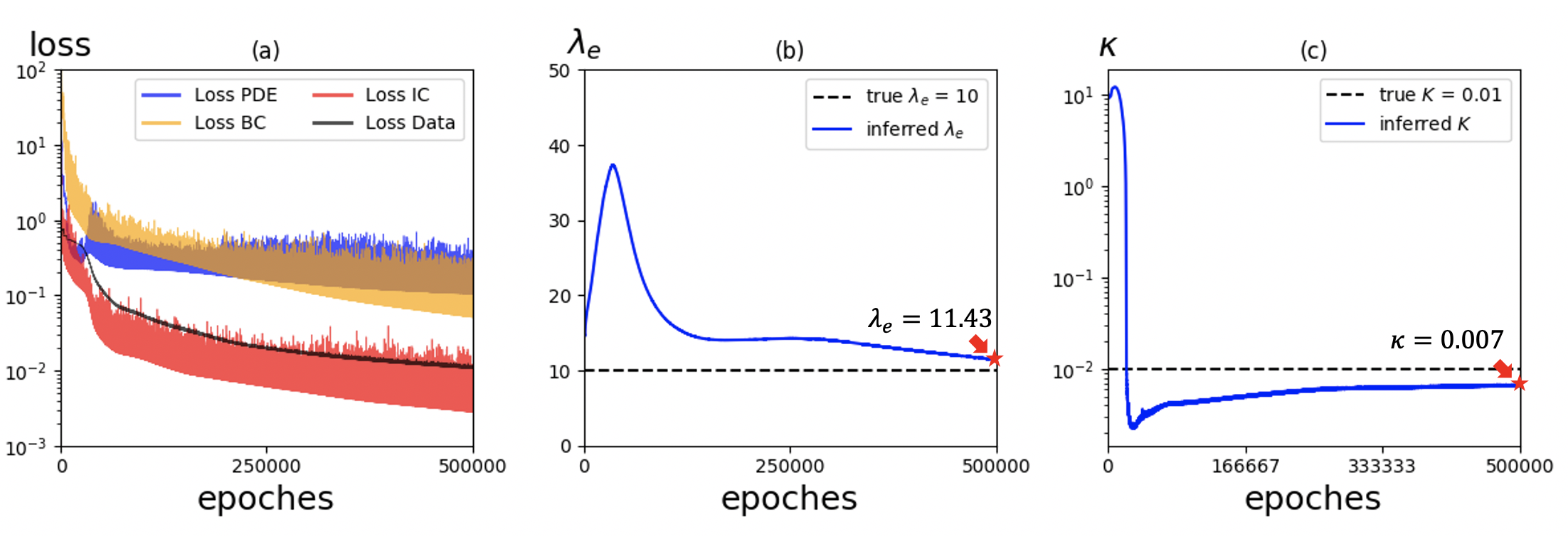}
    \caption{\textbf{History of (a) network losses and inference for (b, c) $\lambda_{e}$ and $\kappa$ against the number of training epochs for the biofilm problem.}}
    \label{fig:lmd_his}
\end{figure}

\begin{figure}
    \centering
    \includegraphics[width=1\textwidth]{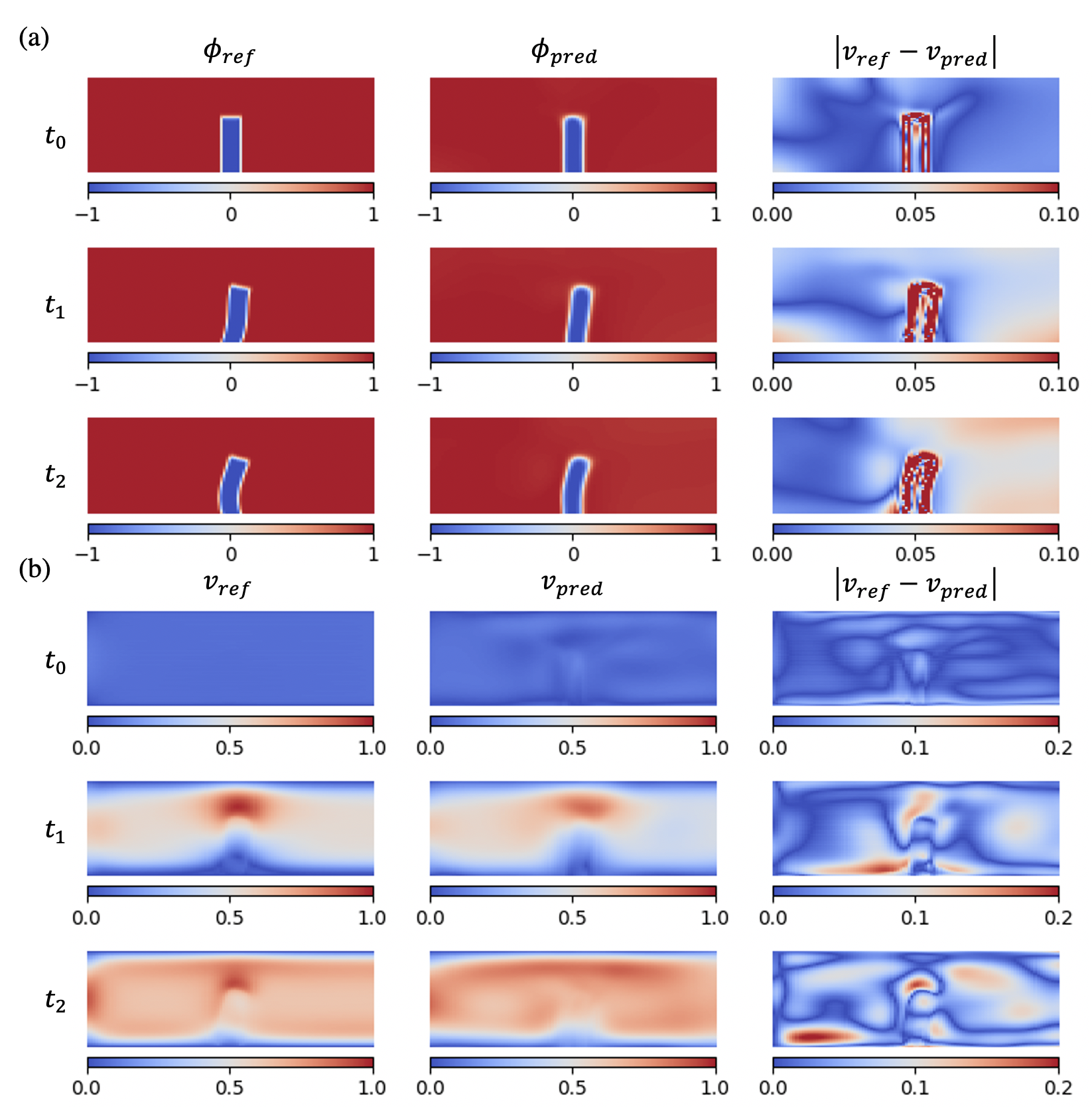}
    \caption{\textbf{2D flow past a visco-elastic biofilm.} With $\lambda_{e}$ and $\kappa$ unknown at the same time, we sample 30,000 points with inner points phase field data and pressure at boundaries scattered in the spatio-temporal region as the training data for parameters inference and field regression. Representative snapshots (at $t_{0}$ = 0.02, $t_{1}$ = 0.44, and $t_{2}$ = 0.86) of the reference (a) phase field and (b) velocity fields are shown against the predicted phase and velocity from the model. The first column shows the reference fields from simulation, the second shows the predicted results from PINNs, and the third column shows the absolute value of the difference between the references and the model predictions. }
    \label{fig:lmd_comp}
\end{figure}

In the first test, with $\lambda_{e}$ and $\kappa$ unknown, we aim to infer the parameters and recover the whole field from the sampled 30,000 training data. The training data contains only phase field data on the domain and pressure measurements at the boundaries. Fig. \ref{fig:lmd_his}(a) shows the history of the training losses where the Loss PDE is the largest among all losses and the loss for data measurements and initial conditions are the lowest. The inferred values for the two unknown parameters are plotted in plot (b) and (c), indicating that the predicted $\lambda_{e}$ and $\kappa$ converge towards the actual values at 11.43 and 0.007 as compared to the true value of $\lambda_{e}=10$ and $\kappa=0.01$. To show the regressed fields, we present the comparisons of the reference data and model predictions for phase and velocity fields at time $t_{0} = 0.02$, $t_{1} = 0.44$, and $t_{3} = 0.86$ in Fig. \ref{fig:lmd_comp}. The first column of Fig. \ref{fig:lmd_comp} shows the actual distribution of phase and velocity fields from the reference data, and the second column shows the regressed fields from the PINN model. In plot (a), we observe overall good phase predictions from the model in the second column with some minor smoothing effects around the sharp interface of the biofilm. Plot (b) presents the comparisons for the velocity field, and we observe that the fluid is forced to pass from the top of the biofilm, with local velocity acceleration because of the impermeability of the biofilm. The model predictions $v_{pred}$ show the capability of the PINN for capturing such effect and regressing the velocity field. The absolute error for the velocity field is generally below 10\% with larger differences at the flow restricted area and close to the bottom boundary. We summarize the mean relative $L_{2}$ error in Table \ref{table:biofilm_err}; the mean error for the first half time window is relatively smaller than the full time. 

\begin{table}
\centering
\begin{adjustbox}{width=1\textwidth}
 \begin{tabular}{c | c | c | c | c | c } 
 \hline
 & $\epsilon_{v}$ & $\epsilon_{\phi}$ & $\epsilon_{p}$ & $\epsilon_{\psi_{1}}$ & $\epsilon_{\psi_{2}}$\\ 
 \hline
 \hline
 Full& $8.104\times10^{-2}$ & $2.580\times 10^{-2}$ & $3.888\times 10^{-4}$ & $9.337\times 10^{-3}$ & $2.018\times 10^{-3}$   \\
 \hline
 Half& $5.057\times10^{-2}$ & $2.115\times 10^{-2}$ & $2.001\times 10^{-4}$ & $5.406\times 10^{-3}$ & $8.561\times 10^{-4}$   \\
 \hline
\end{tabular}
\end{adjustbox}
\caption{\textbf{Summary of the mean relative $L_{2}$ error for the biofilm problem over half and full time window.}}
\label{table:biofilm_err}
\end{table}

\begin{figure}
    \centering
    \includegraphics[width=1\textwidth]{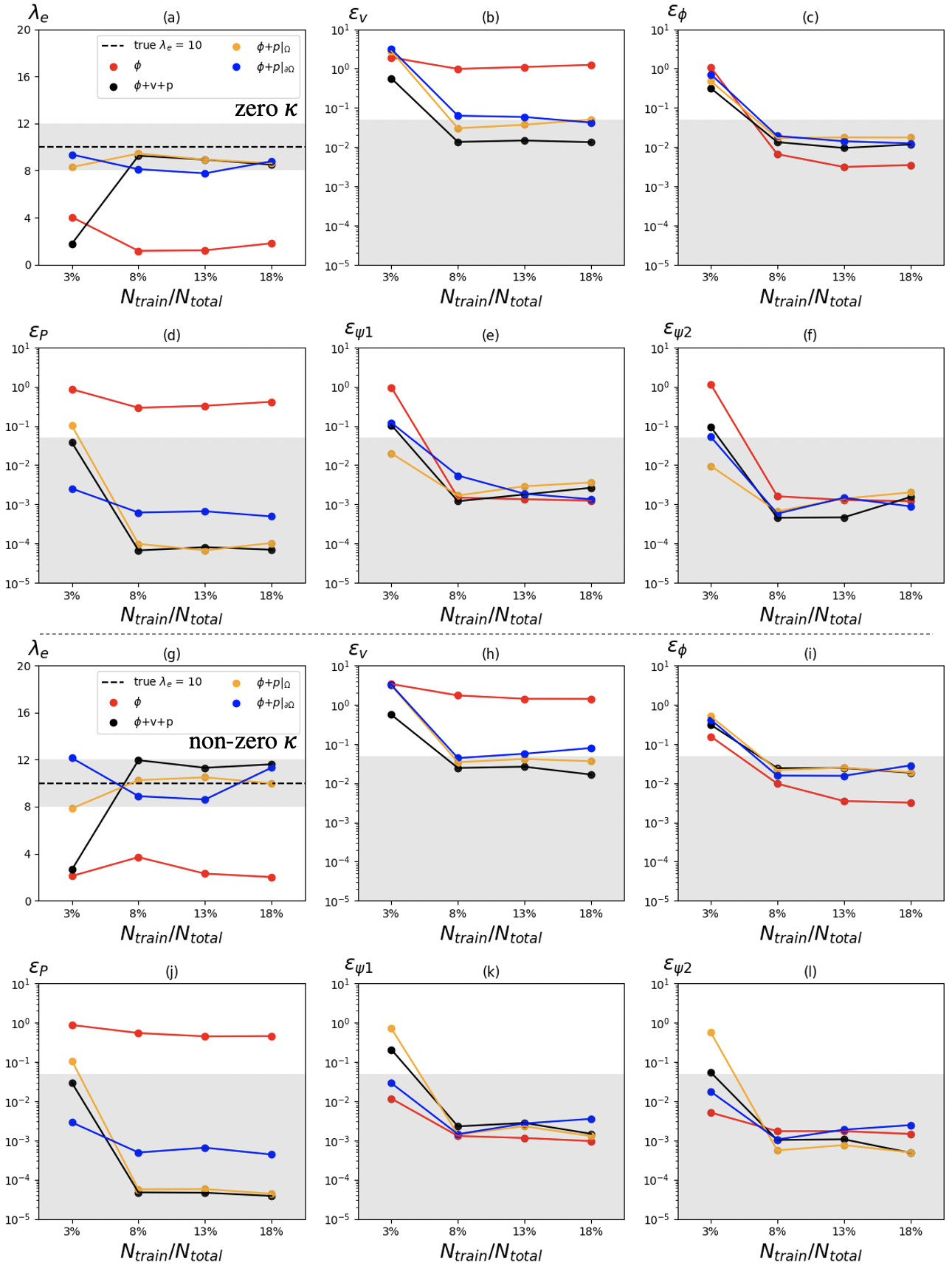}
    \caption{\textbf{Effect of the number of training points on the prediction value of $\lambda_{e}$ with $\kappa$ value as known.} (a, g) Predicted $\lambda_{e}$ versus the number of training points for case with zero $\kappa$ and non-zero $\kappa$. (b-f, h-l) The mean relative error $L_{2}$ versus the number of training points for each field when $\kappa$ = 0 and $\kappa$ = 0.001. We also plotted the inference results and $L_{2}$ error given various training data sources such as $\phi+u+p$, $\phi+p$, and only with $\phi$. For the one with pressure information and phase field, inner ($\Omega$) and boundary ($\partial$) pressure measurements are used respectively to train the neural network. We shaded areas with lower than 5\% of error. }
    \label{fig:lmd_num}
\end{figure}

Furthermore, we present the results as assessments to the inference ability with different numbers and types of data. The number of training points is indicated by the ratio of the number of training points and the number of total points in Fig. \ref{fig:lmd_num}. We set $\kappa$ as a known parameter to avoid its interference in the PINN ability to predict the visco-elastic modulus. The first two rows in Fig. \ref{fig:lmd_num} present the inferred $\lambda_{e}$ and mean of relative $L_{2}$ error for each field for the biofilm problem. The third and fourth rows show the same results for $\kappa$ = 0.01. The inferred modulus shows a convergence to the true value with an increasing amount of data used in training. Hence, the more training points and more data sources we use, the more accurate are the results we obtain. Moreover, we shaded the 20\% error region between $\lambda_{e}$ = 8 and 12 in (a, g) which includes most of the points for yellow, black, and blue lines. As a contrast, the errors in parameter inference exhibit a poor performance if only the phase field data are employed to train the neural network. Additionally, we also compare the change of the errors for each variable with the increasing number of training points in (b-f) and (h-l), and almost all points are fallen below or on the verge of the 5\% error region shaded with gray color except the training with too small training data or modality. These results demonstrate that it is sufficient to infer $\lambda_{e}$ with a limited amount of data from the phase field and pressure measurements at the boundaries. 

\begin{figure}
    \centering
    \includegraphics[width=1\textwidth]{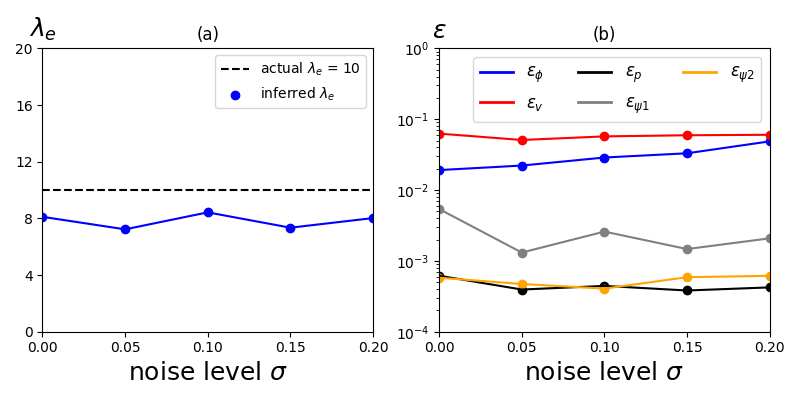}
    \caption{\textbf{2D flow past a visco-elastic biofilm trained with noisy measurements with $\lambda_{e}$ unknown.} (a) The inferred $\lambda_{e}$. (c) presents the mean relative $L_{2}$ errors between the model predictions and the references for each field. Noise is added to the phase field with the noise level ranged from 0 to 20\%. Here, 10,000 data points are scattered in the spatio-temporal domain with 8,000 points sampled around the biofilm.}
    \label{fig:lmd_noise}
\end{figure}

Finally, we investigate the inference ability on $\lambda_{e}$ of the model to noisy measurements. We repeat a set of similar noisy tests as that in permeability inference. Results in Fig. \ref{fig:lmd_noise} are obtained with 10,000 noisy phase and boundary pressure data with the maximum noise level at 20\%. The noise level is similar to that defined in equation \ref{equ:noise} where we denote the percentage with respect to the noise variance. Fig. \ref{fig:lmd_noise} (a) shows the inferred visco-elastic modulus at various noise levels. The inferred value is around 8 with some minor oscillations. Plot (b) illustrates that all the errors exhibit no or very little increase with the increasing intensity in noise level. Hence, we conclude that the PINN shows good robustness to noisy measurements given these findings.

\section{Discussion} 
\label{sec:discussion}
In this paper, we demonstrate the potential of PINNs to infer biological material properties, i.e., permeability and visco-elastic modulus, from relatively limited data. Such modeling leverages the recent advances in deep learning algorithms for scientific machine computing by penalizing the Cahn-Hilliard and Naiver-Stokes equations, which provides a mathematical description for thrombus deformation. Our findings agree well with permeability reference values in a wide range, i.e., from $10^{-3}$ to $10^{2}$, and the model predictions match with the simulation results from the high-order spectral/\textit{hp} element method. In particular, only based on the phase field distribution, the PINN model inferred the value of permeability for a thrombus in a channel, suggesting a potential approach to directly estimate material properties from imaging data. For the inference of visco-elastic modulus, we show that it will be sufficient to make the inference given that the phase field data along with some pressure measurements at boundaries are employed as input to the model. We also demonstrated the robustness of model inference with noisy measurements for both parameters. In addition, we successfully use PINNs to address a thrombus with a space-dependent permeability, i.e., different permeabilities at the core and shell layer. 

In general, we have demonstrated that PINNs can regress the entire fields and unknown parameters given only partial measurements. This provides a novel and viable way to incorporate data from imaging techniques and multi-modality data to train physics-informed deep learning models for field regression and non-invasive parameter inference in biomedical systems. A possible future improvement will be to employ instead of the governing PDEs, the Gibbs energy functional that is minimized to derive the PDEs; this may be advantageous as lower-order derivatives as well as a smaller number of equations is involved. One possible limitation of the current study for biomedical applications is that our model requires the system must have explicit governing equations, whereas most of biological processes are too complicated to be modeled by PDEs; however, in ongoing studies, we have seen that even approximate models can be employed to obtain reasonable results. Also, high dimensionality and lack of boundary/initial information could deteriorate the overall accuracy of the model predictions. In ongoing work, we are extending our research by exploring the possibility of using imperfect PDE constraints and also noise-filtering techniques for realistic imaging data. Furthermore, although PINNs are less data-hungry compared to traditional data-driven models, the amount of data used in training, from an experimental point of view, is still quite intensive given the small dimensions and limited spatio-temporal resolution of imaging. In addition, material properties for a real thrombus are heterogeneous and anisotropic, which could pose additional challenges on the inference of their values. These are important issues that can only be addressed using real multi-modality and multi-fidelity data, and we plan to extend the framework developed herein in future work.

\section*{Acknowledgment}
The work is supported by grant U01 HL142518 of National Institute of Health.

\section*{Appendix}
\begin{center}
\textit{A1. Fields Comparison for biofilm with different $\lambda_{e}$}
\end{center}

\begin{figure}
    \centering
    \includegraphics[width=1\textwidth]{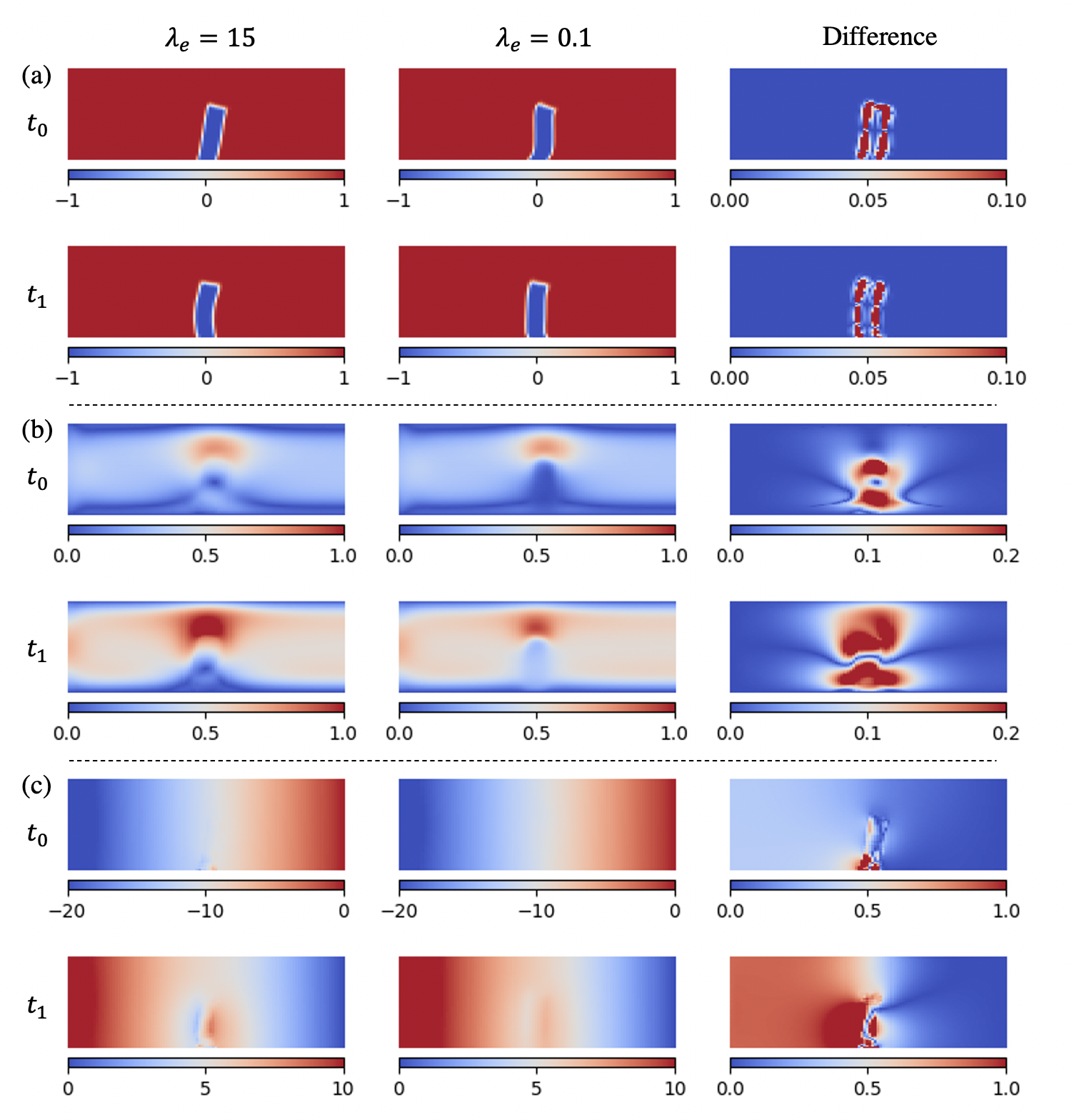}
    \caption{\textbf{2D flow past a visco-elastic biofilm for $\lambda_{e}$ = 15 and 0.1.} The first two lines show the phase field different at $t_{0} = 0.44$ and $t_{1} = 0.86$. The middle two lines and the last two lines show the comparison of velocity and pressure at $t_{0}$ and $t_{1}$. The last column presents the absolute difference between fields data when $\lambda_{e}=15$ and 0.1.}
    \label{fig:append_lmd_comp}
\end{figure}

We present fields comparison when $\lambda_{e} = 0.1$ and 15 for the biofilm problem. In Fig. \ref{fig:append_lmd_comp} (a), we observe small differences between the deformation shape of the biofilm at the same time where most of the differences appear at the interface. Plot (b) shows the comparison between the velocity field where we observe an acceleration area on the top of the biofilm. However, the main inconsistency is caused by the velocity inside the biofilm. In plot (c) we show the pressure field comparison and we observe an overall similar pressure distribution for $\lambda_{e}$ at different values with minor difference close to the center. Overall, the results in Fig. \ref{fig:append_lmd_comp} demonstrate that the difference for $\lambda_{e}$ at 0.1 and 15 are relatively similar, posing a difficulty on the inverse inference to the unknown parameters. 

\begin{center}
\textit{A2. Supplementary results for the inference of $\lambda_{e}$}
\end{center}

\begin{figure}
    \centering
    \includegraphics[width=1\textwidth]{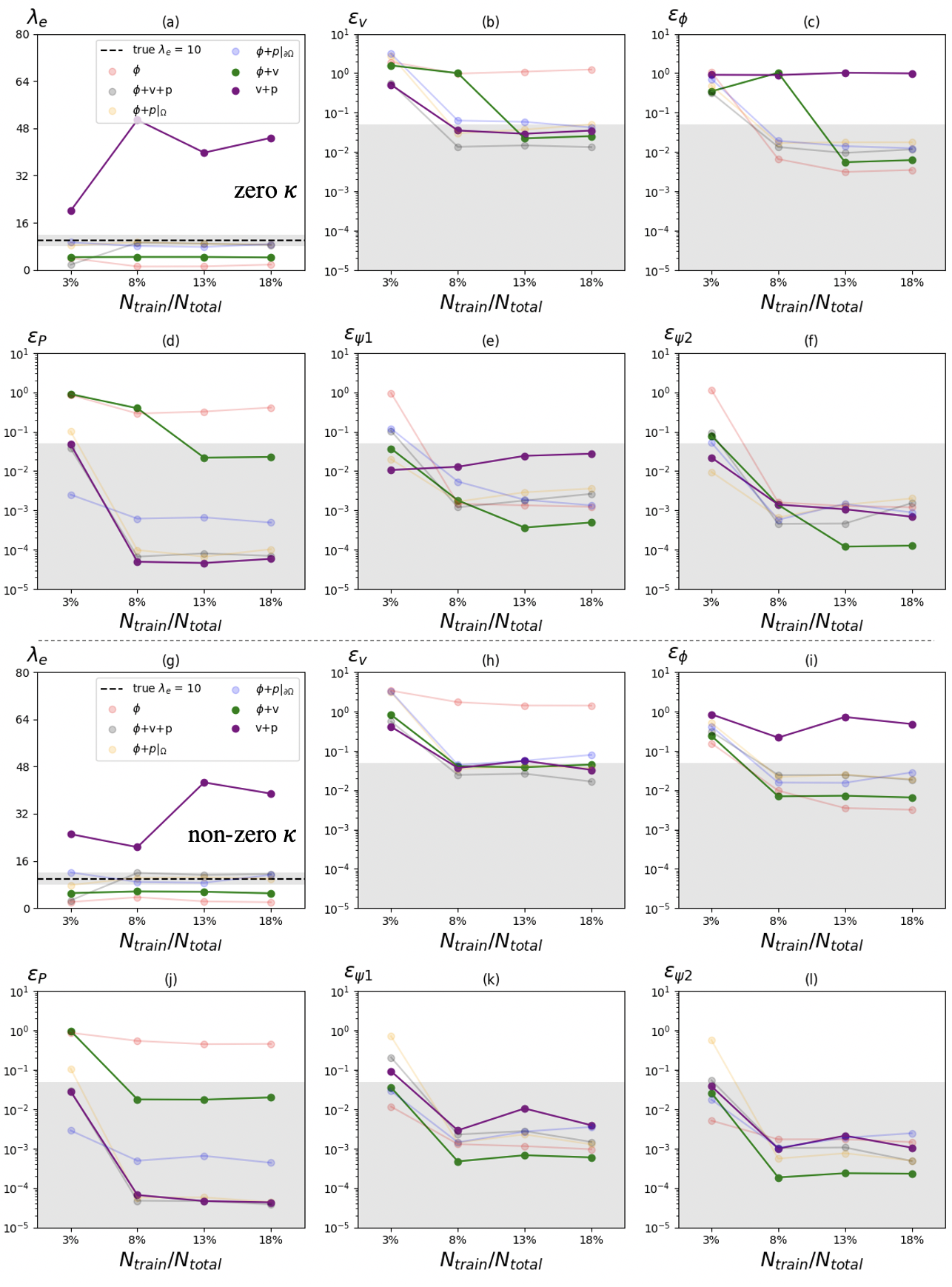}
    \caption{\textbf{Results for the biofilm case with additional results.} We train the PINN model with $p+u$ and $\phi+u$ as comparisons, and we plot the inference results and errors on the top of Fig. \ref{fig:lmd_num} with opaque lines indicating the new results. The shaded areas indicate that the error is lower than 5\%. }
    \label{fig:append_lmd_num}
\end{figure}

In this section, we present two more additional tests result from the network trained with $\phi+u$ (\textcolor{green}{green line}) and $u+p$ (\textcolor{purple}{purple line}) in Fig. \ref{fig:append_lmd_num} on the top of Fig. \ref{fig:lmd_num}. The inferred parameter value on the purple line has the largest error among all lines, and the phase field error ended at the order of 1, indicating converged training results from the PINN model by only using information from velocity and pressure fields. The green line shows the results from training with $\phi$ and $u$. While the error in (e-i) shows a satisfactory agreement with the actual data, the inferred parameter value still cannot match the true value very well. We use these two tests as a supplementary proof to show the importance of pressure data on the inference of $\lambda_{e}$.

\bibliographystyle{abme}
\bibliography{reference}

\end{document}